\documentclass{elsart}
\usepackage{latexsym}
\usepackage{amssymb}
\usepackage[dvips]{graphicx}

\newcommand{\beq}[0]{\begin{equation}}
\newcommand{\eeq}[0]{\end{equation}}

\newcommand{\mc}[1]{\mathcal{#1}}

\begin{document}
\newlength{\caheight}
\setlength{\caheight}{12pt}
\multiply\caheight by 7
\newlength{\secondpar}
\setlength{\secondpar}{\hsize}
\divide\secondpar by 3
\newlength{\firstpar}
\setlength{\firstpar}{\secondpar}
\multiply\firstpar by 2

\begin{frontmatter}
\vskip 48pt
\title{Pricing Derivatives by Path Integral \\
and Neural Networks}

\author[1,2]{Guido Montagna}\author[1,2]{ Marco Morelli}\author[2,1]
{ and Oreste Nicrosini}
\address[1]{Dipartimento di Fisica Nucleare e Teorica, Universit\`a di Pavia\\
Via A. Bassi 6, 27100, Pavia, Italy}
\address[2]{Istituto Nazionale di Fisica Nucleare, sezione di Pavia\\
Via A. Bassi 6, 27100, Pavia, Italy}
\author[3]{Paolo Amato}\author[3]{ and Marco Farina}
\address[3]{ST Microelectronics\\ Softcomputing/Nanoorganics/Silicon Optics/Micromachining Corporate R\&D\\
Via C. Olivetti, 2, Agrate Brianza (Milano), Italy}

\begin{abstract}
Recent progress in the development of efficient
computational algorithms to price 
financial derivatives 
is summarized. A first algorithm is
based on a path integral approach to option pricing, while
a second algorithm makes use of a neural network parameterization
of option prices. The accuracy of the two methods is established
from comparisons with the results of the standard procedures
used in quantitative finance.\\
\end{abstract}
\begin{keyword}
Econophysics; Option Pricing; Path Integral; Neural Networks.\\
{\sc pacs}: 02.50.Ey - 05.10.Gg - 87.18.Sn
\end{keyword}

\end{frontmatter}
\section{Introduction}
\label{intro}
The development of efficient computational algorithms to price and
hedge financial derivatives is a particularly lively topic of the
research work in quantitative finance and econophysics.

In the classical theory of option pricing~\cite{BS} by Black and Scholes,
and Merton (BSM), the price of a financial derivative is
given by a deterministic partial differential equation. In principle,
by solving this equation with the appropriate boundary conditions,
the value of the derivative of interest can be obtained. In practice,
the BSM equation can be analytically solved only in the most simple case
of a so-called European option. Actually, if we
consider other financial derivatives,
which are commonly traded in real markets and allow
anticipated exercise or depend on the history of the
underlying asset, analytical solutions do not exist and numerical
techniques are necessary. As discussed in the literature~\cite{Hu},
the standard numerical procedures are binomial trees,
the method of finite differences and the Monte Carlo simulation of
random walks. Since binomial trees and finite difference
methods are difficult to apply when a detailed control of the paths
is required, the Monte Carlo simulation is the
mostly used method to price path-dependent options.
However, Monte Carlo is known to be time consuming
if precise predictions are required and appropriate variance reduction
techniques must be introduced to save CPU time~\cite{Hu}.

This paper summarizes recent progress in the development of novel
computational methods to price options, as alternatives to traditional
numerical procedures used in finance. A first algorithm is based
on a path integral approach to option pricing, as described in detail
in \cite{MNM}.
A second algorithm relies upon neural networks and
represents work in progress \cite{ip}.

\section{The path integral algorithm}
\label{path}

The path integral method is an integral formulation of the
dynamics of a stochastic process~\cite{W,RT}. It allows to
compute the transition probability associated to
a given stochastic process. We consider, for definiteness, as
stochastic model for the time evolution
of the asset price $S$ the standard BSM Brownian motion,
driven by the stochastic differential equation \cite{BS,Hu}
$d(\ln S)=A dt+ \sigma dw$,
where $\sigma$ is the volatility, $A\doteq\left(\mu-\sigma^{2}/2\right)$,
$\mu$ is the drift parameter and $w$ is a Wiener process.
The probability associated to the above process for
a transition from $z_i = \ln S_i$ to $z_f = \ln S_f$ over a time
interval $\Delta t$ is given by
\beq
 p(z_f|z_i)=\frac{1}{\sqrt{2\pi \sigma^{2} \Delta t}}
 \exp{\left\{-\frac{[z_f-(z_i+A \Delta t)]^{2}}{2\sigma^{2}
 \Delta t}\right\}} .
 \label{eq:gauss}
\eeq
If we consider a finite-time interval [$t',t''$] and we apply a
time slicing in terms of $n+1$ subintervals of length $\Delta t\doteq
(t''-t')/n+1$, the finite-time transition probability
 can be expressed as a convolution of the form \cite{MNM}
\begin{eqnarray}
\!\!\!\!\!\!\!\!\!\int_{-\infty}^{+\infty}\!\!\!\cdots\int_{-\infty}^{+\infty}\!\!\!\!dz_{1}\cdots
 dz_{n} \frac{1}{\sqrt{(2\pi \sigma^{2}\Delta t)^{n+1}}}
 \exp{\left\{-\frac{1}{2\sigma^{2}\Delta t}\sum_{k=1}^{n+1}\left[
 z_{k}-(z_{k-1}+A\Delta t)\right]^{2}\right\}}. \,
 \label{eq:pd}
\end{eqnarray}
Once the transition probability is known, the price of an option
can be computed as the conditional
expectation value of a given functional of the stochastic process,
involving integrals of the form~\cite{MNM,RT}
\beq
 E[\mc{O}_{i}|S_{i-1}]=\int dz_{i}
 p(z_{i}|z_{i-1})\mc{O}_{i}(e^{z_{i}}) .
\label{eq:mi}
 \eeq
For example, for an European call option at the maturity
$T$ the quantity of interest will be ${\rm max}\,\{S_T-X,0\}$, $X$ being the strike price.

\subsection{Transition probability}
\label{tp}

By means of appropriate
manipulations of the integrand entering eq.~(\ref{eq:pd}),
it is possible to
obtain an expression for the transition probability suitable
for an efficient numerical implementation~\cite{MNM}.
If we define $y_{k}\doteq z_{k}-kA\Delta t$, $k=1,\ldots,n$, we
can rewrite eq.~(\ref{eq:pd}) as
 \beq
\int_{-\infty}^{+\infty}\cdots\int_{-\infty}^{+\infty}dy_{1}\cdots
 dy_{n} \frac{1}{\sqrt{(2\pi \sigma^{2}\Delta t)^{n+1}}}
 \cdot\exp{\left\{-\frac{1}{2\sigma^{2}\Delta t}\sum_{k=1}^{n+1}
 [y_{k}-y_{k-1}]^{2}\right\}},
 \label{eq:qq}
 \eeq
 in order to get rid of the drift parameter.
A convenient quadratic form can be extracted from the argument of the exponential function,
 to arrive, after some algebra, at the following
 formula for the probability distribution \cite{MNM}
 \beq
\!\!\!\!\!\!\!\!\int_{-\infty}^{+\infty}\!\!\cdots\!\!\int_{-\infty}^{+\infty}\prod_{i=1}^{n}dh_{i}
 \frac{1}{\sqrt{2\pi \sigma^{2}\Delta t \mathrm{det}(M)}}
 \exp{ \left \{ -\frac{1}{2\sigma^{2}\Delta t} \left[
 y_{0}^{2}+y_{n+1}^{2}+ \sum_{i=1}^{n}
 \frac{(y_{0}O_{1i}+y_{n+1}O_{ni})^{2}}{m_{i}} \right ] \right \}}
 \label{eq:main}
 \eeq
 where we have introduced new
variables $h_{i}$ obeying the relation
 \beq
 dh_{i}\doteq\sqrt{ \frac{m_{i}}{2\pi \sigma^{2}\Delta t}}
 \exp{ \left\{ -\frac{m_{i}}{2 \sigma^{2}\Delta t}
 \left[w_{i}-\frac{(y_{0}O_{1i}+y_{n+1}O_{ni})}
 {m_{i}}\right]^{2}\right\}}dw_{i} .
 \eeq
 The gaussian variables $h_{i}$ can be extracted
from a normal distribution
with mean $(y_{0}O_{1i}+y_{n+1}O_{ni})^{2}/m_{i}$ and variance
$\sigma^{2}\Delta t/m_{i}$, where $m_i$ are the eigenvalues 
of a real, symmetric and tridiagonal matrix $M$ and
$O_{ij}$ are the matrix elements of the
orthogonal matrix $O$ which diagonalizes $M$, with $w_i = O_{ij} y_j$.
The probability distribution, as given by eq.~(\ref{eq:main}), is
  an integral whose kernel is a constant function and can be
very efficiently and precisely computed as shown in \cite{MNM}.

\subsection{Option pricing}
\label{pathi}

In the case of options with possibility
of anticipated exercise before the expiration date,
it is necessary to check at any time $t_i$, and
 for any value of the stock price $S_i$,
 whether early exercise is more convenient with respect to
 holding the option for a future time. Therefore, the value
of the option at a time slice $i$ is given by
\beq
 \mc{O}_{i}(S_{i})=\max\left\{\mc{O}_{i}^{Y}(S_{i}),e^{-r\Delta t}
 E[\mc{O}_{i+1}|S_{i}]\right\} ,
 \eeq
where $E[\mc{O}_{i+1}|S_{i}]$ is the expectation value of $\mc{O}$
at time $i+1$ under
the hypothesis of having  the price $S_{i}$ at the time
$t_{i}$, $\mc{O}_{i}^{Y}$ is the option value in the
case of anticipated exercise and $r$ is the risk-free interest.
To keep under
control the computational complexity,
it is mandatory to
limit the number of points for the integral
evaluation of $E[\mc{O}_{i+1}|S_{i}]$. To this end, we can create a grid of
possible prices, according to the dynamics of the
stochastic process. Starting from $z_{0}$, we
evaluate the expectation value
$E[\mc{O}_{1}|S_{0}]$ with  $p=2m+1,m\in\mathbb{N}$ values of
$z_{1}$ centered on the mean value
$E[z_{1}]=z_{0}+A\Delta t$ and differing of a
quantity of the order of $\sigma\sqrt{\Delta t}$
 \[
 z_{1}^{j}\doteq z_{0}+A\Delta
 t+j\sigma\sqrt{\Delta t}, \quad \quad \ j=-m,\ldots,+m.
 \]
We can then evaluate each expectation value
$E[\mc{O}_{2}|z_{1}^{j}]$ obtained from each one of the $z_{1}$'s
created above with $p$ values for $z_{2}$ centered around the mean
value
 \[
 E[z_{2}|z_{1}^{j}]=z_{1}^{j}+A\Delta t=z_{0}+2A\Delta
 t+j\sigma\sqrt{\Delta t}.
 \]
Iterating the procedure until the maturity, we create a
deterministic grid of points
such that, at a given time $t_{i}$, there are $(p-1)i +1$ values of
$z_{i}$, in agreement with a request of linear growth.

In the particular case of an American option, the possibility of exercise
at {\it any} time up to the expiration date
allows to develop
a specific semi-analytical procedure, which
is precise and very fast \cite{MNM}. In the limit $\Delta t\to 0$ and
$\sigma\to 0$, the
transition probability of eq.~(\ref{eq:gauss}) approximates a
delta function. This means that, apart from volatility effects,
the price $z_{i}$ at time $t_i$ will have
a value remarkably close to the expected value $\bar{z}\doteq
z_{i-1}+A\Delta t$, given by the drift. This suggests
to evaluate integrals
as in eq.~(\ref{eq:mi}) by performing a Taylor expansion of the kernel function
$\mc{O}_{i}(e^{z})$ around the expected value
$\bar{z}$, to arrive, after inserting the analytical expression
(\ref{eq:gauss}), at the following semi-analytical approximation
\beq
 E[\mc{O}_{i}|S_{i-1}]=\mc{O}_{i} (\bar{z})+
 \frac{\sigma^{2}}{2}\Delta t
 \mc{O}_{i}''(\bar{z})
 +\ldots,
 \label{eq:sa}
 \eeq
where the second derivative $\mc{O}_{i}''$ can be estimated numerically.
It is worth noticing that each expectation value
$E[\mc{O}_{i}|S_{i-1}]$ can be now computed once
only $p=3$ points at each slice are known, in order to
evaluate the second derivative,
provided $\Delta t$ is taken sufficiently small.

\subsection{Numerical results and comparisons}
\label{numerical}

To test the algorithm, we
present results for the particular case of the price of an American option
in the BSM model in Tab.~1, as obtained with
the calculation of the transition probability and
the grid technique described above
(path integral 1) and with the semi-analytical approximation
of eq. (\ref{eq:sa}) (path integral 2). As can be seen from Tab.~1, there is generally a good agreement
of our path integral results with those known in the
literature~\cite{RT} and obtained by means of binomial trees,
the finite difference
method and the Green Function
Deterministic Numerical Method (GFDNM) \cite{RT}. It is worth noticing that
our results in the path integral 1 require only a few seconds on
a PentiumIII 500MhZ PC, while the CPU time is negligible for path integral 2.
 Further numerical results for option prices and greek letters can be
found in \cite{MNM}.

\begin{table}
 \caption{Price of an American put option in the BSM
 model for the parameters $t=0$ year, $T=0.5$ year, $r=0.1$,
 $\sigma=0.4$, $X = 10$, as a function different stock prices $S_0$.
 The path integral 1 is
 performed with  200 time slices and $p=13$ integration points, while
 path integral 2 is obtained with 300 time slices and $p=3$.}
 \begin{tabular}{c c c c c c}
 \hline
 \hline $S_{0}$ &  finite difference & binomial tree & GFDNM & path integral 1 & path integral 2 \\
 \hline
 \hline
 6.0 & 4.00 & 4.00 & 4.00 & 4.00 & 4.00\\
 8.0 & 2.095 & 2.096 & 2.093 &  2.095 & 2.095\\
 10.0 & 0.921 & 0.920 & 0.922 &  0.922 & 0.922\\
 12.0 & 0.362 & 0.365 & 0.364 &  0.362 & 0.362\\
 14.0 & 0.132 & 0.133 & 0.133 &  0.132 & 0.132\\
 \hline
 \hline
 \end{tabular}
 \end{table}

\section{Neural networks}
\label{nn}

Because the pricing formulae are given by non-linear functions and neural networks are
known to be well suited to approximate non-linear relations, it is conceivable to
use the fast path integral algorithm above summarized to generate grids of option
prices over which to train learning networks. The goal is to construct neural networks
able to price and hedge derivatives with a sufficient degree of accuracy to be of
practical use.

Using the Neural Networks Toolbox of MatLab, we developed Radial
Basis Functions networks, since it is known that this kind of
networks have the best approximation properties for
real-valued continuous functions~\cite{HLP}. For such networks, the
activation functions are gaussians of variable width (a parameter
called spread). We
trained the networks using as training data the prices for
European and American options obtained with the path integral
algorithm. We generated several training grids of different
dimensionality, starting from grids depending on one and two
parameters
to arrive at
multi-parametric grids. After the training phase, we tested the generalization
performances of the networks, comparing the predictions of the networks for
input parameters not present in the training sample with the results of the benchmark.
In general, we observed that the performances of the networks 
strongly depend on the value
of the spread parameter and the number of data used in 
the training samples, as expected.

\begin{figure}
 \begin{center}
 \vspace{3mm}
 \includegraphics[scale=0.6]{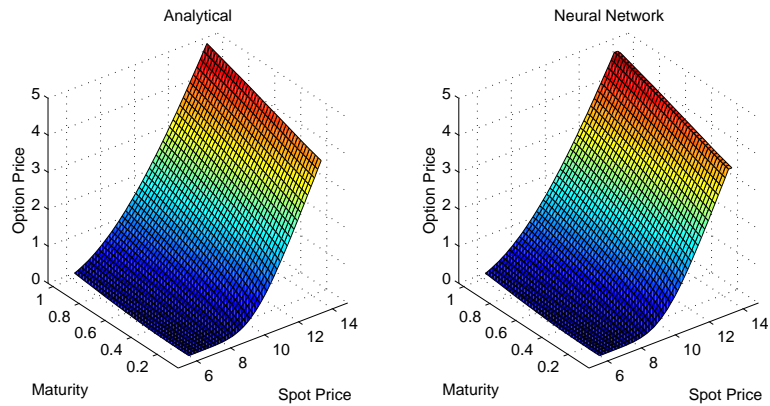}
 \caption{Comparison for an European call option between the
 analytical results (left)
 and the neural network predictions (right), as a function of
 spot price and maturity. The BSM parameters are $X = 10$,
 $\sigma=0.4$, $r = 0.1$. $20 \times 20$ data points are used for
 training and $40 \times 40$ for generalization.}
 \label{fig:nn}
 \end{center}
 \end{figure}

An example of the results obtained is shown in Fig.~\ref{fig:nn}, for a two-parameter case of an
European call option in the BSM model, as a function of spot price and maturity. The exact analytical values
(left) differ on average from the neural networks predictions (right) well
below the 1\% level.
More details and further numerical results will be presented
elsewhere~\cite{ip}.

\section{Conclusions}
\label{conclusion}

Computational algorithms based on the path integral approach
to stochastic processes and neural networks can be successfully applied to the
problem of option pricing in financial analysis.

The path integral can provide fast and accurate
predictions for a large class of financial derivatives
with path-dependent and early exercise features, by means of a careful evaluation of the
transition probability and a suitable choice of the integration points needed to evaluate
the quantities of financial interest. The computational cost
of the algorithm is competitive with the most efficient strategies
used in finance. Also neural networks are a powerful and flexible tool for option pricing.
With them it is possible to parameterize functions
emerging from financial models in a very efficient and flexible way.
After an appropriate training, neural networks can predict
option prices with good accuracy, even in the case of 
multi-parametric dependence. 

The natural development of the present work
concerns the application of our computational methods
of option pricing to more realistic models of financial dynamics,
beyond the gaussian approximation~\cite{lb}.

\vskip 8pt\noindent
{\bf Acknowledgments} \\
One of us (G.~M.) thanks the organizers, and in particular H.E. Stanley, for the
kind invitation.

\end{document}